\documentclass[conference,a4paper]{IEEEtran}
\usepackage{graphicx}
\usepackage{amsmath}
\usepackage{amssymb}
\usepackage{mathrsfs}
\usepackage{stfloats}
\usepackage{dsfont}
\usepackage{setspace}
\usepackage{epstopdf}
\usepackage{cite}
\usepackage{balance}
\usepackage{xcolor}
\usepackage{booktabs}
\usepackage{pifont}

\usepackage{accents}

\newtheorem{lemma}{Lemma}

\renewcommand{\QED}{\QEDopen}
\begin{document}
%
\title{
Coverage Probability and Spectral Efficiency Analysis of Multi-Gateway Downlink LoRa Networks
} 
\author{
\authorblockN{Lam-Thanh Tu, Abbas Bradai and Yannis Pousset}
\authorblockN{ Institute XLIM, University of Poitiers, France}
 \authorblockN{ e-mail: \{lam.thanh.tu, abbas.bradai, yannis.pousset\}@univ-poitiers.fr }

%
}
%
%
\maketitle
\vspace{-5cm}
\begin{abstract}
	The system-level performance of multi-gateway downlink long-range (LoRa) networks is investigated in the present paper. 
	Specifically, we first compute the active probability of a channel and the selection probability of an active end-device (ED) in the closed-form expressions. We then derive the coverage probability (Pcov) and the area spectral efficiency (ASE) under the impact of the capture effects and different spreading factor (SF) allocation schemes. 
	Our findings show that both the Pcov and the ASE of the considered networks can be enhanced significantly by increasing both the duty cycle and the transmit power.
	Finally, Monte-Carlo simulations are provided to verify the accuracy of the proposed mathematical frameworks.
\end{abstract}
%
\begin{keywords}
LPWAN, LoRa Networks, Stochastic Geometry, System-Level Analysis, Spectral Efficiency.
\end{keywords}
%
%
\vspace{-0.35cm}
\section{Introduction} \label{Introduction}
\vspace{-0.15cm}
It is expected that the number of connected devices via wireless technology will be over 20 billion by 2022 \cite{EricReport:01} and this number is likely to increase significantly to realize the Internet of Things (IoT).
Additionally, there are many IoTs devices located in remote areas that are not able to connect to the power grid to replenish its battery.
Thus, short-range technologies such as ZigBee, Bluetooth Low Energy (BLE), and other technologies that belong to the family of the IEEE 802.15.4 standards are not suitable for connecting such long-range transmission, low power consumption devices. 
Recently, low power wide area networks (LPWAN) technology has emerged as a promising solution to fulfill these requirements \cite{Claire:01}. 
Among all available LPWAN technologies, long-range (LoRa) has received many attractions from both academia and industry thanks to its advantageous properties, i.e., long-range transmissions, energy-efficient, and wide coverage areas.
LoRa is primarily designed for uplink traffic, from end-devices (ED) to gateway. Hence, it is no surprise that most of the existing literature concentrates on best-effort applications, showing the scenarios with solely unconfirmed messages and leaving the downlink traffic although it is vital as well. For example, the acknowledgment (ACK) messages sent to EDs to confirm the data reception, the \textit{join-accept} message to accept the \textit{join-request} of EDs to join the networks, and the control messages to adjust the transmission parameters such as LinkADRReq, LinkCheckAns. etc \cite{LoRa:DL_constraint:01}. Moreover, all mission-critical applications require a reliable data transport system where the ACK messages are unavoidable \cite{Maalel:LoRa:01}.

The performance of uplink LoRa networks was studied extensively in \cite{LoRa:analysis:seminal:01,SG_Matern:01,Beltramelli:ALOHA:01,Reynders:faire-sf:01,Thanh:WCL:01}. In \cite{LoRa:analysis:seminal:01}, the system-level coverage probability (Pcov) of ED was examined where EDs were modelled by a homogeneous Poisson point process (PPP). 
The performance of both Pcov and area spectral efficiency (ASE) were investigated in \cite{SG_Matern:01} where EDs are distributed according to the  Matern cluster process. 
Compared with PPP, the Matern cluster process provides more accurately the characteristics of the EDs. It, however, leads to mathematical intractability thus, numerical computations are employed for computing these metrics. The Pcov and the energy efficiency of the uplink single gateway LoRa networks were addressed under different MAC protocols, i.e., pure ALOHA, slotted ALOHA, and CSMA \cite{Beltramelli:ALOHA:01}. The findings show that the slotted ALOHA attains the highest performance among all considered protocols. 
In \cite{Reynders:faire-sf:01}, the authors proposed a novel spreading factor (SF) allocation scheme that ensures a fair collision probability among all spreading factors. This scheme is superior to the popular distance-based allocation scheme in terms of the packet error rate.
Although many papers have been published on the uplink performance of LoRa networks, the performance of another direction is fairly attractive. Indeed, the performance of downlink LoRa networks were studied in \cite{LoRa:DL_constraint:01,Centenaro:01,Ron:01}. More precisely, Valentina \textit{et. al.} in \cite{LoRa:DL_constraint:01} addressed the downlink scalability of LoRa via computed-based simulations rather than rigorous mathematical frameworks. Their findings show that multi-gateway is beneficial for downlink LoRa networks and overcomes the bottleneck of downlink networks, the gateway, due to the constraint of the duty cycle. The authors in \cite{Centenaro:01} investigated the impact of downlink feedback of the uplink traffic via computer-based simulations.
In \cite{Ron:01}, authors examined the average waiting time of the downlink frame of class B devices. 
%

Building upon the aforementioned works the present paper investigates the system-level performance of multi-gateway downlink LoRa networks where both gateways and EDs are distributed according to a homogeneous PPP. More precisely, our main contributions are summarized as follows: i) the impact of the duty cycle on the availability of the channel is taken into consideration; ii) both co- and inter-SF interference (capture effects) are taken into account at the receiver;
iii) the active probability of an arbitrary channel and the selection probability of an active ED are computed in the closed-form expressions; iv) Pcov and ASE under two SF allocation schemes, namely the fair-collision and random schemes, as well as co- and inter-SF interference are derived; and
v) Monte-Carlo simulations are supplied to verify the accuracy of the proposed frameworks.
%
%
\vspace{-0.35cm}
\section{System Model} \label{sec:System}
%
%
\vspace{-0.15cm}
\subsection{LoRa Networks Modeling}
\vspace{-0.15cm}
Let us consider a downlink multi-gateway LoRa networks where gateways and EDs follow by two independent homogeneous Poisson distribution processes denoted by ${\Psi _{{\rm{GW}}}}, {\Psi _{{\rm{ED}}}}$ with densities ${\lambda_{\rm{GW}}} \ll {\lambda _{\rm{ED}}}$, respectively.
The PPP modelling is adopted due to its mathematical tractability and is a lower bound of the practical measurement \cite{Andrews:SG:01}. Thus, it can be applied to any practical scenario.
We employ a slotted ALOHA protocol and all messages are taken place within one time-slot \cite{Beltramelli:ALOHA:01}. 
The slotted ALOHA is chosen since it achieved the best performance among all available protocols as proven in \cite{Beltramelli:ALOHA:01}.
Additionally, each gateway has $N_{\text{Ch}} = 8$ channels for downlink transmission and all channels are regulated by the duty cycle $\rho$ ranging from 0.1\% to 10\% depending on the regulator \cite{Tech_Report_ETSI:01}. 
Additionally, we adopt an equal power allocation among all channels, the transmit power of each channel is then computed as $P_{{\rm{tx}}}^{{\rm{Ch}}} = {{P_{{\rm{tot}}}}} / {{N_{{\rm{Ch}}}}}$ where $P_{\text{tot}}$ is the total transmit power.
The developed mathematical frameworks are for the typical ED denoted by $\text{ED}_0$ which is located at the origin (Slivnyak theorem \cite[Th. 1.4.5]{Baccelli:book:01}) and the gateway which serves the $\text{ED}_0$ is referred to $\text{GW}_0$. The cell association criterion is provided in Section \ref{SubSec:Cell_Asso}. 
Interference created from other technologies that operate at the same industrial, scientific, and medical (ISM) band does not take into account as it is a typical case in the literature \cite{LoRa:analysis:seminal:01,Thanh:LoRa_ICC:01}. 
\vspace{-0.2cm}
\subsection{Channel Modelling}
Considering a generic link from an arbitrary gateway to a generic end-device, it is subjected to both small-scale fading and large-scale path-loss. The impact of the shadowing is not taken into account since it can implicitly be examined by appropriately scaling the density of GWs \cite{Thanh:EE_Cellular:01}.
\subsubsection{Small-scale fading} 
Let us denote $h_n$ as the small-scale fading of an arbitrary link from a gateway to ED and is modelled by a Rayleigh distribution.
The channel gain $|h_{n}|^2$, as a result, follows an exponential distribution with mean $\Omega$. For simplicity, we assume that $\Omega_n = 1, \forall n$.
\subsubsection{Large-scale path-loss}
Let us consider a generic link from a gateway to an ED the large-scale path-loss is defined as 
\vspace{-0.35cm}
\begin{align}
    \label{Sec:SYSTEMS:eq:02}
    L_{n} = l \left( r_{n} \right) = K_{0} r_{n}^{\beta}.
\end{align}
Here $r_{n}$ is the distance from the gateway to the ED; $K_{0} = \left( 4\pi {f_c} / c \right)^2$ and $\beta > 2$ are the path-loss constant and the path-loss exponent, respectively.
$f_c$ is the carrier frequency (in Hz) and $c = 3 \times 10^8$ (in meters per second) is the speed of light.
\subsection{Cell Association Criterion} \label{SubSec:Cell_Asso}
ED is served by the gateway which has the smallest path-loss. Let us denote $\Psi _{{\rm{GW}}}^{\left( {\rm{A}} \right)}$ as the set of \textit{available} gateways. 
A gateway is considered as \textit{available} if it has at least one available channel at each time slot.
The serving GW, $\text{GW}_0$, is then formulated as follows:
\vspace{-0.2cm}
\begin{align} \label{Eq:SecII:cell_asso}
    {\rm{G}}{{\rm{W}}_0} = \mathop {\text{argmin} }\limits_{n \in \Psi _{{\rm{GW}}}^{\left( {\rm{A}} \right)}} \left\{ {{L_n}} \right\}.
\end{align}
It is noted that $\Psi _{{\rm{GW}}}^{\left( {\rm{A}} \right)}$ is also a homogeneous PPP with density $\lambda _{{\rm{GW}}}^{\left( {\rm{A}} \right)} = \mu {\lambda _{{\rm{GW}}}}$ according to the thinning property of PPP \cite{Baccelli:book:01}. $\mu  = 1 - {\left( {1 - \rho } \right)^{{N_{{\rm{Ch}}}}}}$ is the probability having one available channel of a gateway.
\subsection{Spreading Factor Allocation}
\subsubsection{Fair-collision scheme}
We consider the fair-collision SF allocation scheme that was proposed in \cite{Reynders:faire-sf:01}. The considered SF allocation method guarantees a fair-collision probability among all available SFs. 
Additionally, it is proven that the selected scheme is far better than the popular distance-based allocation scheme \cite{Reynders:faire-sf:01}. 
The probability that an ED is assigned to SF$k$ from its serving gateway is then computed as
\begin{align} \label{Eq:SF_allocation:01} 
    {p_k^{\text{fa}}} = \left( k / 2^k \right) / \sum\nolimits_{i = 7}^{12} \left( i / 2^i \right),
    \;\;\; k \in \left\{ 7, \ldots, 12 \right\}.
\end{align}
As a consequence, the density of EDs utilized SF$k$ under this scheme is calculated as $\lambda_{\text{ED,fa}}^{k} = p_k^{\text{fa}} \lambda_{\text{ED}}$.
\subsubsection{Random scheme}
Differently under a random scheme, each ED is randomly assigned to an arbitrary SF from its serving gateway. 
Thus, the probability of EDs utilized SF$k$ is $p_k^{\text{ra}} = 1 / 6, \forall k$ and the density of EDs utilized SF$k$ is given as $\lambda_{\text{ED,ra}}^k = p_k^{\text{ra}} \lambda_{\text{ED}}$.
\subsection{Interference Modeling}
In the present work, we consider the aggregate interference from all active gateways apart from the serving GW, $\text{GW}_0$. A gateway is defined as active if it does not belong to one of the two following cases: i) there is no ED in its coverage area; ii) EDs are presented in its coverage area, nevertheless, no downlink transmission happens. 
The second condition is originated from the main purpose of LoRa networks.
Particularly, LoRa is designed for uplink transmission, thus downlink traffic is very limited compared to its counterpart and occurs solely when EDs request acknowledges or control messages from the gateway. 
An ED demanding information is called active ED and the set of active EDs is denoted by $\Psi _{{\rm{ED}}}^{\left( {\rm{A}} \right)}$ that follows by a homogeneous PPP with density $\lambda _{{\rm{ED}}}^{\left( {\rm{A}} \right)} = \theta \lambda _{{\rm{ED}}}$ due to the thinning property of PPP, $\theta$ is the active probability of EDs.
The adopted interference modelling is certainly more accurate than one considering the strongest interference \cite{LoRa:analysis:seminal:01,Thanh:LoRa_ICC:01}, especially when a number of interferers go up without bound.
Moreover, we also consider a practical scenario where the orthogonality between SFs is imperfect thus, the receiver suffers from both co- and inter-SF interference instead of only co-SF interference like in \cite{LoRa:analysis:seminal:01,Thanh:LoRa_ICC:01}.
\subsection{Load Modeling}
Let us denote ${N_{{\rm{ED}}}} \in \mathbb{N}$ and $\tau \in \left\{ 1, \ldots, N_{\text{Ch}} = 8 \right\}$ are the number of active EDs and the number of available channels in a generic cell at one time-slot. If ${N_{{\rm{ED}}}} \ge \tau$ then the serving gateway will randomly select $\tau$ active EDs to serve. It is noted that due to the hardware constraint, the gateway can serve merely one ED per channel in one time-slot. Thus, there is no intra-cell interference and the inter-cell interference is present on a channel basis. 
On the contrary, if ${N_{{\rm{ED}}}} < \tau$ all ${N_{{\rm{ED}}}}$ active EDs will be served by its serving BS. 
The consider random scheduling at each transmission instance ensures that all active EDs associated with a gateway are scheduled for transmission in the long term manner.
%
%
\vspace{-0.2cm}
\section{Preliminary results}
\vspace{-0.1cm}
In this section, we summarize some intermediate results which are important for computing the two considered metrics, namely the coverage probability and the area spectral efficiency. 
More precisely, the active probability of a generic channel of an available gateway is employed to identify the set of active interferer when computing the SIR condition of the Pcov, i.e., the probability that the intended signal is greater than the sum of active gateways transmitting at the same channel. On the other hand, the selection probability of an active ED is used to compute the ASE which are formulated in \eqref{Eq:ASE:definition:01}.
\vspace{-0.2cm}
\subsection{Active probability of a channel of an available gateway}
The active probability of a channel of an available gateway refers to the probability that a channel is randomly chosen among all available channels to serve an active ED and is provided by Lemma \ref{Lem:ActiveProb_Channel} as follows:
\begin{lemma} \label{Lem:ActiveProb_Channel}
    Considering $\mathcal{A} = \lambda _{{\rm{ED}}}^{\left( {\rm{A}} \right)}/\lambda _{{\rm{GW}}}^{\left( {\rm{A}} \right)}$ the active probability of a channel denoted by $P_{\text{Act}}$ is calculated as
    \begin{align} \label{Eq:Pact:01}
        {P_{{\rm{Act}}}} =& \sum\limits_{i = 1}^{{N_{\rm{Ch}}}} {} \left( {1 - \left( {\frac{{{N_{{\rm{Ch}}}} - i}}{{{N_{{\rm{Ch}}}}}} + \frac{i}{{{N_{{\rm{Ch}}}}}}\sum\limits_{k = 0}^{i - 1} {} \left( {1 - \frac{k}{i}} \right){\mathcal{T}_1}\left( k \! \right)} \! \right)} \! \right) 
        \nonumber\\
        & \;\;\;\;\;\;\;\;\; \times \mathcal{V} \left( {{N_{{\rm{Ch}}}},i,\rho } \right) / \mu
        \nonumber \\
        \mathcal{T}_1 \left( k \right) =& \frac{{{{\left( {3.5} \right)}^{3.5}}\Gamma \left( {k + 3.5} \right){\mathcal{A}^k}}}{{\Gamma \left( {3.5} \right)k!{{\left( {\mathcal{A} + 3.5} \right)}^{k + 3.5}}}},
    \end{align}
    where
    $\mathcal{V} \left( {{N_{{\rm{Ch}}}},i,\rho } \right) = \left( {\begin{array}{*{20}{c}}
{{N_{{\rm{Ch}}}}}\\
i
\end{array}} \right){\rho ^i}{\left( {1 - \rho } \right)^{{N_{{\rm{Ch}}}} - i}}$ is the probability having $i$ available channel out of $N_{\rm{Ch}}$ channels, $\mathcal{T}_1 \left( k \right)$ is the probability having $k$ active EDs in a generic cell \cite{Yu:VoronoiCell:01}, and $\Gamma \left( . \right)$ is the Gamma function.

\begin{proof}
        The proof can be derived in two main steps. 
        In the first step, we apply the total probability theorem for the number of available channels conditioned on at least one available channel.
        In the second step, given the number of available channels, we compute the active probability of a channel via the inactive probability and employ the total probability theorem for the number of active EDs, $N_{\text{ED}}$, associated with the gateway.
\end{proof}
\end{lemma}
\vspace{-0.1cm}
\subsection{Selection probability of an active ED}
The selection probability of an active ED refers to the probability that an active ED is scheduled for transmission in an available channel. 
This probability accounts for the fact that the number of active EDs is greater than the number of available channels. Thus, some active EDs can be served and the others are blocked. 
The selection probability denoted by ${P_{\rm{Sel}}}$ is computed in Lemma \ref{Lem:ProbSel} as follows:
\begin{lemma} \label{Lem:ProbSel}
    Considering $\mathcal{A} = \lambda _{{\rm{ED}}}^{\left( {\rm{A}} \right)}/\lambda _{{\rm{GW}}}^{\left( {\rm{A}} \right)}$ the selection probability of an active ED can be computed as follows:
    \begin{align}  \label{Eq:PSel:01}
        {P_{{\rm{Sel}}}} =& \sum\limits_{i = 1}^{{N_{\rm{Ch}}}} {} \frac{\mathcal{V}\left( {{N_{{\rm{Ch}}}},i,\rho } \right)}{\mu }\left( {1 - {\mathcal{B}_1 \left( i \right)}\left( {{\mathcal{B}_2 \left( i \right)} - {\mathcal{B}_3 \left( i \right)}} \right)} \right),
    \end{align}
where    
    \begin{align}  \label{Eq:PSel:01a}
        {\mathcal{B}_1}\left( i \right) =& \frac{{{{\left( {3.5} \right)}^{3.5}}\Gamma \left( {i + 4.5} \right){\mathcal{A}^i}}}{{\Gamma \left( {3.5} \right){{\left( {\mathcal{A} + 3.5} \right)}^{i + 4.5}}}}
        \nonumber \\
        {\mathcal{B}_2}\left( i \right) =& \frac{{{\;_2}{F_1}\left( {1,i + 4.5,i + 1, \mathcal{A} /\left( {\mathcal{A} + 3.5} \right)} \right)}}{{\Gamma \left( {i + 1} \right)}}
        \nonumber \\
        {\mathcal{B}_3}\left( i \right) =& \frac{ i {{\;_2}{F_1}\left( {1,i + 4.5,i + 2, \mathcal{A} /\left( {\mathcal{A} + 3.5} \right)} \right)}}{{\Gamma \left( {i + 1} \right)}},
    \end{align}
    where ${\;_2}{F_1}\left( . \right)$ is the Gaussian hypergeometric function.

    \begin{proof}
        The proof is also derived in two main steps and the first step is similar to Lemma \ref{Lem:ActiveProb_Channel}.
        In the second step, the selection probability of an active ED given $i$ available channels denoted by ${P_{{\rm{Sel}}}}\left( i \right)$ is then formulated as 
        \begin{align}
            {P_{{\rm{Sel}}}}\left( i \right) = \sum\limits_{k = 0}^{i - 1} {} \Pr \left\{ {N_{{\rm{ED}}}^{'} = k} \right\} + \sum\limits_{k = i}^\infty  \frac{i 
            \Pr \left\{ {N_{{\rm{ED}}}^{'} = k} \right\}
            }{{k + 1}},
        \end{align}
        where $N_{{\rm{ED}}}^{'}$ is the number of other active EDs
in the coverage area of a gateway conditioned on an
ED has already associated to this gateway; $\Pr \left\{ . \right\}$ is the probability operator; $\Pr \left\{ {N_{{\rm{ED}}}^{'} = k} \right\}$ is held with the help of \cite[Lemma 3]{Yu:VoronoiCell:01} and by formulating the summations in terms of the Gaussian hypergeometric function we finish the proof.
    \end{proof}
\end{lemma} 
%
\vspace{-0.3cm}
\section{Performance Analysis} \label{Sec:PerAnalysis}
\vspace{-0.1cm}
The present analysis concentrates on two vital metrics of LoRa networks, i.e., the Pcov and the ASE.
The Pcov particularly focuses on the performance of LoRa networks from the viewpoint of an ED while the ASE is from the network’s point
of view. Thus, the examination of two metrics permits a holistic assessment of the network performance.
\vspace{-0.3cm}
\subsection{Coverage Probability}
\vspace{-0.1cm}
An active ED of SF$k$ is considered as successfully received information from its serving gateway, provided that both signal-to-noise ratio (SNR) and signal-to-interference ratio (SIR) conditions are satisfied.
\subsubsection{SIR condition analysis}
The SIR condition is the probability that the ratio of the intended signal utilized SF$k$ to the aggregate interference from all active GWs transmitting on the same channel utilized SF$\widetilde{k}$ is greater than a pre-defined threshold.
More precisely, we investigate two aggregate interference cases: i) only co-SF interference; ii) both co- and inter-SF interference. 
The former corresponds to the perfect orthogonality between signals with different SFs, while the latter considers imperfect orthogonality between these signals.
\paragraph{Co-SF interference analysis}
The SIR condition under co-SF interference of an SF$k$ ED under SF allocation scheme $s$ is formulated as follows:
\begin{align}  \label{Eq:PcovSIR:Intra:01}
    P_{{\mathop{\rm cov}} }^{{\rm{SIR,co}}}\left( {k,s} \right) = \Pr \left\{ {{\rm{SI}}{{\rm{R}}_{k,k}^s} \ge {\Delta _{k,k}}} \right\} = {P_{{\rm{SIR}}}}\left( {k,k,s} \right),
\end{align}
where ${\Delta _{k,\widetilde k}}$ (in dB), $k, \widetilde{k} \in \left\{ {7, \ldots, 12} \right\}$ is the $k$-row, $\widetilde{k}$-column of matrix $\Delta$ in \eqref{Eq:Table:01} and accounts for the interference rejection threshold that depends on the SF$k$ of the intended signal and SF$\widetilde{k}$ of the interferer \cite{Croce:01};  ${\rm{SI}} {{\rm{R}}_{k,\widetilde k}^s} = \left(  {\left| {{h_0}} \right|}^2 / L_0\right)/{I_{\widetilde k,s}}$; ${I_{\widetilde k,s}} = \sum\nolimits_{z \in \psi _{_{\widetilde k,s}}^{\left( \rm{I} \right)\backslash 0}} {} \left(  {\left| {{h_z}} \right|}^2 / L_z\right) \textbf{1}\left( {{L_z} > {L_0}} \right)$ is the aggregate interference from all active gateways apart from the GW$_0$ transmitting on the same channel as the typical link and utilizing SF$\widetilde{k} \in \left\{ 7,\ldots, 12 \right\}$ under $s$ SF allocation method; $\textbf{1}\left( . \right)$ is the indicator function; the condition $\textbf{1}\left( {{L_z} > {L_0}} \right)$ ensures that all interferer lie outside the circle whose radius is the distance from GW$_0$ to ED$_0$ and center at ED$_0$;
${{{\left| {{h_0}} \right|}^2}}$, ${{{\left| {{h_z}} \right|}^2}}$ and ${{L_0}}$, ${{L_z}}$ are the small-scale fading and large-scale path-loss of the serving GW$_0$ and interferer GW$_z$ to ED$_0$;
$\psi _{_{\widetilde k,s}}^{\left( \rm{I} \right)\backslash 0}$ is the set of interference gateways that serves ED of SF$\widetilde{k}$ under SF allocation scheme $s$ and is approximated as a homogeneous PPP with density $\lambda _{{\rm{GW}}}^{\left( {\rm{I}} \right),\widetilde k,s} = p_{\widetilde k}^s{P_{{\rm{Act}}}}\lambda _{{\rm{GW}}}^{\left( {\rm{A}} \right)}$.
{\small
\begin{align} \label{Eq:Table:01}
& \;\;\;\;\;\;\;\;\;\;\;\begin{array}{*{20}{c}}
\textrm{SF}_7&\textrm{SF}_8&\textrm{SF}_9&\textrm{SF}_{10}&\textrm{SF}_{11}&\textrm{SF}_{12}
\end{array}
\nonumber \\
\!\! \Delta  \! =& \!\! 
\begin{array}{*{20}{c}}
\textrm{SF}_7\\
\textrm{SF}_8\\
\textrm{SF}_9\\
\textrm{SF}_{10}\\
\textrm{SF}_{11}\\
\textrm{SF}_{12}
\end{array} \!\!\! \left[ {\begin{array}{*{20}{c}}
	1&{ - 8}&{ - 9}&{ - 9}&{ - 9}&{ - 9}\\
	{ - 11}&1&{ - 11}&{ - 12}&{ - 13}&{ - 13}\\
	{ - 15}&{ - 13}&1&{ - 13}&{ - 14}&{ - 15}\\
	{ - 19}&{ - 18}&{ - 17}&1&{ - 17}&{ - 18}\\
	{ - 22}&{ - 22}&{ - 21}&{ - 20}&1&{ - 20}\\
	{ - 25}&{ - 25}&{ - 25}&{ - 24}&{ - 23}&1
	\end{array}} \!\!\! \right].
\end{align}
}
${P_{{\rm{SIR}}}}\left( {k,k,s} \right)$ in \eqref{Eq:PcovSIR:Intra:01} is computed in the closed-form expression with the help of Lemma \ref{Lem:PcovSIR} and is given as
%
%
\vspace{-0.2cm}
\begin{lemma} \label{Lem:PcovSIR}
    Let us define a shorthand $\Theta \left( x \right) = {\;_2}{F_1}\left( {1, - \delta ,1 - \delta , - x} \right) - 1$, $\delta =2 / \beta$, the $P_{S{\rm{IR}}}\left( {k,\widetilde k}, s \right)$ is then computed as follows
\begin{align} \label{Eq:lem1:PcovSIR:01}
    P_{S{\rm{IR}}}\left( {k,\widetilde k}, s \right) = 
    {\left( {1 + p_{\widetilde k}^s{P_{{\rm{Act}}}}\Theta \left( {{\Delta _{k,\widetilde k}}} \right)} \right)^{ - 1}}.
\end{align}
\begin{proof}
    See Appendix \ref{Appendix:PcovSIR:01}.
\end{proof}
\end{lemma}
By employing Lemma \ref{Lem:PcovSIR}, $P_{{\mathop{\rm cov}} }^{{\rm{SIR,co}}}\left( {k,s} \right)$ in \eqref{Eq:PcovSIR:Intra:01} is then evaluated as
\begin{align}
    P_{{\mathop{\rm cov}} }^{{\rm{SIR,co}}}\left( {k,s} \right) = 
    {\left( {1 + p_{k}^s{P_{{\rm{Act}}}}\Theta \left( {{\Delta _{k,k}}} \right)} \right)^{ - 1}}.
\end{align}
%
%
\vspace{-0.2cm}
\paragraph{Both co- and inter-SF interference analysis}
The SIR condition under both co- and inter-SF interference of ED utilized SF$k$ with $s$ SF allocation method is given as
\begin{align}  \label{Eq:PcovSIR:Both:01}
    P_{{\mathop{\rm cov}} }^{{\rm{SIR,bo}}}\left( {k,s} \right) =&
    \prod\nolimits_{\widetilde k = 7}^{12} {\Pr \left\{ {{\rm{SIR}}_{k,\widetilde k}^s \ge {\Delta _{k,\widetilde k}}} \right\}} 
    \nonumber \\
    =& 
    \prod\nolimits_{\widetilde{k}=7}^{12}  {\left( {1 + p_{\widetilde k}^s{P_{{\rm{Act}}}}\Theta \left( {{\Delta _{k, \widetilde{k} }}} \right)} \right)^{ - 1}}.
\end{align}
%
%
%
\subsubsection{SNR condition analysis}
The SNR condition refers to the reception condition is the probability that received signals at the ED of SF$k$ is greater than the quality-of-service (QoS) requirement and is given as
{\small
\begin{align} \label{Eq:PcovSNRcond:01}
    & P_{{\rm{cov}}}^{{\rm{SNR}}}\left( k \right) = \Pr \left\{ {\frac{{P_{{\rm{tx}}}^{{\rm{Ch}}}{{\left| {{h_0}} \right|}^2}}}{{{\sigma ^2}{L_0}}} \ge {\gamma _{{\rm{D}},k}}} \right\} = \delta \pi \lambda _{{\rm{GW}}}^{\left( {\rm{A}} \right)}{\left( {{K_0}} \right)^{ - \delta }}
    \nonumber \\
    & \times
    \int\limits_{x = 0}^\infty  {} {x^{\delta  - 1}}\exp \left( { - \frac{{{\gamma _{{\rm{D}},k}}{\sigma ^2}}}{{P_{{\rm{tx}}}^{{\rm{Ch}}}}}x - \pi \lambda _{{\rm{GW}}}^{\left( {\rm{A}} \right)}{{\left( {\frac{x}{{{K_0}}}} \right)}^\delta }} \right)dx.
\end{align}
}
The integration in \eqref{Eq:PcovSNRcond:01}, unfortunately, is not able to compute in the closed-form expression due to the arbitrary value of the path-loss exponent, i.e., $\beta > 2$. It, however, can be straightforwardly computed by employing numerical methods via some popular software like Matlab or Mathematica. 
${\gamma_{\rm{D}}}$ is the QoS threshold and is a function of the spreading factor $k$, particularly, we have ${\gamma_{\rm{D}}} = \left\{ - 6, - 9, - 12, - 15, - 17.5, - 20 \right\}$ dBm for SF7 to SF12, respectively.
${{\sigma ^2}} = -174 + \text{NF} + 10 \log_{10} \left( \text{Bw} \right)$ (in dBm) is the noise variance of AWGN noise; NF is the noise figure (in dBm) and Bw is the transmission bandwidth (in Hz).

Having obtained the mathematical framework of two conditions, the coverage probability of a SF$k$ ED under SF allocation scheme $s$ and $o \in \left\{ \text{co, bo} \right\}$ interference denoted by ${P_{{\mathop{\rm cov}}}^{o}} \left( k,s \right)$ is computed as 
\begin{align} \label{Eq:Pcov:def:01}
    {P_{{\mathop{\rm cov}}}^{\rm{co}}  } \left( k,s \right) =& P_{{\mathop{\rm cov}} }^{{\rm{SNR}}}\left( k \right)
    {\left( {1 + p_{k}^s{P_{{\rm{Act}}}}\Theta \left( {{\Delta _{k,k}}} \right)} \right)^{ - 1}}
     \\
    {P_{{\mathop{\rm cov}}}^{\rm{bo}}  } \left( k,s \right) =&
    P_{{\mathop{\rm cov}} }^{{\rm{SNR}}}\left( k \right)
    \prod\nolimits_{\widetilde k = 7}^{12}
    {\left( {1 + p_{\widetilde k}^s{P_{{\rm{Act}}}}\Theta \left( {{\Delta _{k, \widetilde{k} }}} \right)} \right)^{\! - 1}}\!.
    \nonumber
\end{align}

\subsection{Area Spectral Efficiency}
The area spectral efficiency (in bit/s/$\text{m}^2$) measures the network information rate per unit area which satisfies the minimum QoS objectives, as imposed by the reliability thresholds $\gamma_{\text{D},k}$ and $\Delta_{k,\widetilde{k}}$. Mathematical speaking, the ASE of the considered networks under $s \in \left\{ \text{fa, ra} \right\}$ SF allocation scheme and $o \in \left\{ \text{co, bo} \right\}$ interference is given by 
\begin{align} \label{Eq:ASE:definition:01}
    {\rm{AS}}{{\rm{E}}_s^{\rm{o}}} = \sum\limits_{k = 7}^{12} {p_k^s\lambda _{{\rm{ED}}}^{\left( {\rm{A}} \right)}} {{\cal R}_k}{P_{{\rm{Sel}}}}{P_{{\rm{cov}}}^{o} }\left( {k,s} \right) = \sum\limits_{k = 7}^{12} {{\rm{AS}}{{\rm{E}}_{k,s}^{o}}} 
\end{align}
where $\mathcal{R}_k = k \left( \text{Bw} / 2^k \right) \left( 4 / (4 + \text{Cr})  \right)$ is the bit rate of SF$k$; $\text{Cr} \in \left\{ 1, \ldots, 4 \right\}$ is the coding rate.
\vspace{-0.2cm}
\section{Numerical Results}
\vspace{-0.1cm}
This section provides numerical results to verify the accuracy of the proposed mathematical frameworks in Section \ref{Sec:PerAnalysis}.
Unless otherwise stated, the following setup is considered: $\beta = 2.9$, $\text{Bw}$ = 125 kHz, NF = 6 dBm, $f_c$ = 868 MHz, $P_\text{tot}$ = 25 dBm, $\text{Cr}$ = 1, $N_{\text{Ch}} = 8$, ${\lambda _{{\rm{GW}}}} = 2 / \text{km}^2$, $\rho = 1\%$, ${\lambda _{{\rm{ED}}}} = 1000 / \text{km}^2$, $\theta = 0.01$.
\begin{figure}[!ht]
    \centering
    \includegraphics[width=0.425\textwidth]{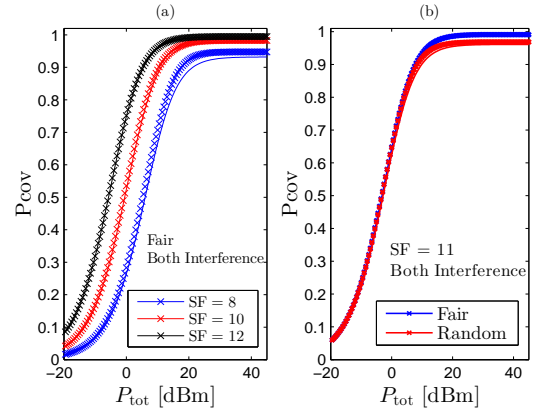}
    \caption{
            Coverage probability versus the total transmit power. Solid lines are plotted by using \eqref{Eq:Pcov:def:01}. Markers are Monte-Carlo simulations. 
				}
    \label{fig:sec:num:vsPtx:01}
\end{figure}

Fig. \ref{fig:sec:num:vsPtx:01} illustrates the coverage probability versus the total transmit power $P_{\text{tot}}$. Firstly, it is evident that the proposed mathematical frameworks are extremely tight compared with the Monte-Carlo simulations. Secondly, turning up the transmit power is beneficial to the Pcov as it is a monotonic increasing function. Thirdly, Fig. \ref{fig:sec:num:vsPtx:01}(a) reveals that under the fair-collision allocation scheme the larger the SF, the better the Pcov owing to the smaller $p_k^{\rm{fa}}$ of the larger $k$.
Additionally, observing Fig. \ref{fig:sec:num:vsPtx:01}(b) we experience that $P_{{\mathop{\rm cov}} }^{{\rm{bo}}} \left( {11,{\rm{fa}}} \right)$ outperforms $P_{{\mathop{\rm cov}} }^{{\rm{bo}}}\left( {11,{\rm{ra}}} \right)$, although the gap is negligible when $P_{\text{tot}}$ is fair and moderate.
\begin{figure}[!ht]
    \centering
    \includegraphics[width=0.42\textwidth]{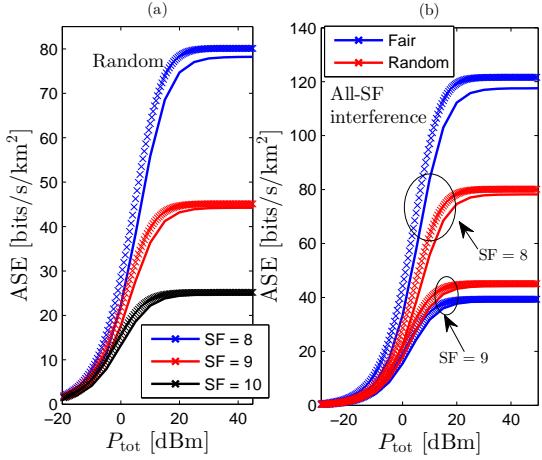}
    \caption{
           Area spectral efficiency versus the total transmit power. Solid lines are plotted by using \eqref{Eq:ASE:definition:01}. Markers are Monte-Carlo simulations.
				}
    \label{fig:sec:num:vsPtx:02}
\end{figure}

Fig. \ref{fig:sec:num:vsPtx:02} sketches the ASE with respect to the $P_{\text{tot}}$. We again experience that the developed framework is close to the exact computer-based simulations. It is apparent that ASE goes up with the transmit power and has a similar trend as the Pcov. The main reason is that ASE in \eqref{Eq:ASE:definition:01} is equal to Pcov multiplying with some factors that are independent of the $P_{\text{tot}}$. 
Besides, the smaller the SF, the higher the bit rate $\mathcal{R}$ is the main reason that increasing SF will improve the ASE.
Relying on the relation of the $p_k^{\rm{fa}}$ to the $p_k^{\rm{ra}}$, the ASE of the fair-collision scheme maybe outperform or underperform compared with its counterpart. Particularly, we have $\text{ASE}_{8,\text{fa}}^{\rm{bo}} > \text{ASE}_{8,\text{ra}}^{\rm{bo}}$ due to $p_8^{\rm{fa}} > p_8^{\rm{ra}}$ and $\text{ASE}_{9,\text{fa}}^{\rm{bo}} < \text{ASE}_{9,\text{ra}}^{\rm{bo}}$ because of $p_9^{\rm{fa}} < p_9^{\rm{ra}}$ as shown in Fig. \ref{fig:sec:num:vsPtx:02}(b). 
\begin{figure}[!ht]
    \centering
    \includegraphics[width=0.42\textwidth]{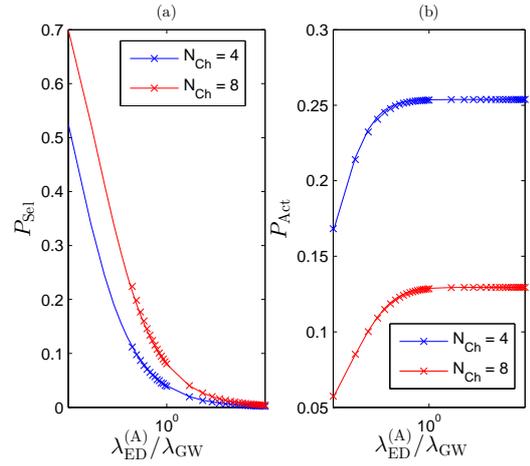}
    \caption{
            $P_{\rm{Sel}}$ (a) and $P_{\rm{Act}}$ (b) versus the ratio of $\lambda _{{\rm{ED}}}^{\left( {\rm{A}} \right)}/{\lambda _{{\rm{GW}}}}$ with various gateway channels. Solid lines are plotted by using \eqref{Eq:Pact:01} and \eqref{Eq:PSel:01}. Markers are Monte-Carlo simulations.
				}
    \label{fig:sec:num:vsRatio:01}
    \vspace{-0.4cm}
\end{figure}

Fig. \ref{fig:sec:num:vsRatio:01} reveals the behavior of $P_{\text{Sel}}$ and $P_{\text{Act}}$ regarding the ratio $\lambda _{{\rm{ED}}}^{\left( {\rm{A}} \right)}/{\lambda _{{\rm{GW}}}}$. We obverse that augmenting the number of channels $N_{\rm{Ch}}$ will ameliorate the $P_{\text{Act}}$ but decrease the $P_{\text{Sel}}$.
Concerning the $P_{\text{Sel}}$, increasing $\lambda _{{\rm{ED}}}^{\left( {\rm{A}} \right)}/{\lambda _{{\rm{GW}}}}$ means that there are more active EDs tagged to each cell. Thus, the probability that an ED is selected by the GW will decline. On the contrary, the more active EDs, the higher the active probability of each channel hence, $P_{\text{Act}}$ keeps increasing with $\lambda _{{\rm{ED}}}^{\left( {\rm{A}} \right)}/{\lambda _{{\rm{GW}}}}$. 
Nevertheless, due to the constraint of the duty cycle $\rho$, the upper bound of the $P_{\text{Act}}$ will never approach one, provided that the number of active ED goes without bound.

\begin{figure}[!ht]
    \centering
    \includegraphics[width=0.42\textwidth]{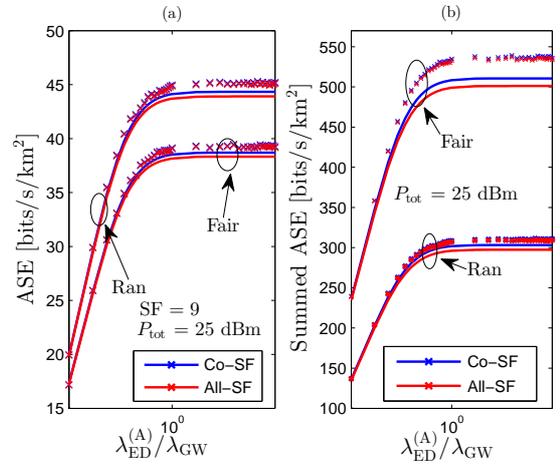}
    \caption{
            Area spectral efficiency versus the ratio of $\lambda _{{\rm{ED}}}^{\left( {\rm{A}} \right)}/{\lambda _{{\rm{GW}}}}$. Solid lines are plotted by using \eqref{Eq:ASE:definition:01}. Markers are Monte-Carlo simulations.
				}
    \label{fig:sec:num:vsRatio:02}
    \vspace{-0.5cm}
\end{figure}

Fig. \ref{fig:sec:num:vsRatio:02} illustrates the ASE of specific SFs (a) and the summed ASE of the whole networks (b) versus the ratio $\lambda _{{\rm{ED}}}^{\left( {\rm{A}} \right)}/{\lambda _{{\rm{GW}}}}$.
The figure unveils that the ASE is monotonically increasing with respect to $\lambda _{{\rm{ED}}}^{\left( {\rm{A}} \right)}/{\lambda _{{\rm{GW}}}}$. 
From Figs. \ref{fig:sec:num:vsPtx:02}(b) and \ref{fig:sec:num:vsRatio:02}(a), we observe that ASE under random assignment scheme can be either better or worse than another scheme. Nonetheless, observing Fig. \ref{fig:sec:num:vsRatio:02}(b), the summed ASE of the whole networks under fair-collision scheme is superior to its counterpart.
Additionally, the ASE suffers only from co-SF interference is always higher than one is subjected to both co- and inter-SF interference. Nevertheless, the gap between the two schemes is negligible.
\begin{figure}[!ht]
    \centering
    \includegraphics[width=0.42\textwidth]{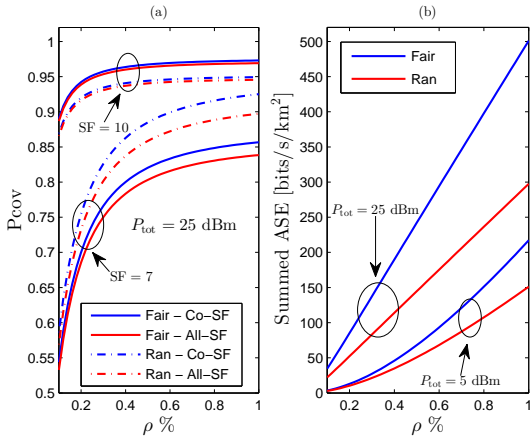}
    \caption{
            Coverage probability (a) and summed ASE (b) versus the duty cycle $\rho$. Solid lines are plotted by using \eqref{Eq:Pcov:def:01} and \eqref{Eq:ASE:definition:01}. 
				}
    \label{fig:sec:num:vsrho:01}
\end{figure}

Fig. \ref{fig:sec:num:vsrho:01} investigates the behavior of Pcov (a) and the summed ASE (b) as a function of the duty cycle $\rho$. This figure highlights the benefits of increasing the duty cycle as both metrics ameliorate significantly with $\rho$. 
Particularly, the summed ASE under fair-collision scheme improves over 10-fold when $\rho$ increases from 0.1\% to 1\% for $P_{\text{tot}}$ = 25 dBm while the improvement under random assignment scheme is also higher than 6-times. Again, Fig. \ref{fig:sec:num:vsrho:01}(b) confirms the superior of fair-collision scheme compared with random assignment one.
Additionally, Fig. \ref{fig:sec:num:vsrho:01}(a) highlights that the gap of the Pcov under co- and all-SF interference is either negligible or significantly relying on the utilized SF. Specifically, $P_{\text{cov}}^{\rm{co}} \left( 7, s\right)$ is higher than $P_{\text{cov}}^{\rm{bo}} \left( 7, s\right)$ nearly 0.1 while $P_{\text{cov}}^{\rm{co}} \left( 10, s\right)$ and $P_{\text{cov}}^{\rm{bo}} \left( 10, s\right)$ is nearly indistinguishable.
%
%
\vspace{-0.3cm}
\section{Conclusion}
\vspace{-0.25cm}
The mathematical frameworks of the downlink multi-gateway LoRa networks were examined in the present paper. More precisely, two important metrics namely, the coverage probability and the area spectral efficiency were computed under both SF fair-collision and random assignment schemes and both co- and inter-SF interference. Our results revealed that improving the duty cycle and the transmit power was important to ameliorate both metrics. 
Numerical results were also provided to clarify the correctness of the proposed mathematical frameworks. 
The paper can be extended in several directions and one of the most important ones is to do the practical experiments for comparing with the proposed mathematical framework. 
%
\vspace{-0.35cm}
\appendix
\subsection{Proof of Eq. \eqref{Eq:lem1:PcovSIR:01} } \label{Appendix:PcovSIR:01}
\vspace{-0.25cm}
In this section, we derive \eqref{Eq:lem1:PcovSIR:01}, let us start with the definition of the SIR condition as follows:
\begin{align} \label{Eq:Apn01:01}
    & P_{S{\rm{IR}}}\left( {k,\widetilde k}, s \right) = \Pr \left\{ {{\left| {{h_0}} \right|}^2} / \left( L_0 I_{\widetilde k,s} \right)  \ge {\Delta _{k,\widetilde k}} \right\}
    \nonumber \\
    & \mathop  = \limits^{\left( a \right)} \int\nolimits_{l = 0}^\infty  {} \int\nolimits_{i = 0}^\infty  \exp \left( { - il{\Delta _{k,\widetilde k}}} \right){f_{{I_{\widetilde k,s}}}}\left( i \right){f_{{L_0}}}\left( l \right)didl
    \nonumber \\
    & \mathop  = \limits^{\left( b \right)} \int\nolimits_{l = 0}^\infty {M_{ {I_{\widetilde k,s}} }}\left( {\left. {l{\Delta _{k,\widetilde k}}} \right|l} \right){f_{{L_0}}}\left( l \right)dl,
\end{align}
where $\left( a \right)$ is held by utilizing the cumulative distribution function (CDF) of the exponential function; $\left( b \right)$ is obtained via employing the definition of the moment generating function (MGF) of ${I_{\widetilde k,s}}$ conditioned on $L_0$. 
Looking at \eqref{Eq:Apn01:01}, in order to solve this integration, we first identify the MGF of ${I_{\widetilde k,s}}$ conditioned on $L_0$ as follows:
\vspace{-0.15cm}
\begin{align} 
    & {M_{{I_{\widetilde k,s}}}}\left( {\left. s \right|l} \right) = \mathbb{E} \left\{ {\exp \left( { - s{I_{\widetilde k,s}}} \right)} \right\}
     \mathop  = \limits^{\left( a \right)} 
    \exp \left(  {\delta \pi \lambda _{{\rm{GW}}}^{\left( {\rm{I}} \right),\widetilde k,s}} / {\left( {{K_0}} \right)}^\delta
    \right.
    \nonumber \\
    & \left. \times
    {\mathbb{E}_{h}}\left\{ {\int\nolimits_{x = l}^\infty  \left( {1 - \exp \left(  - sh/x \right)} \right){x^{\delta  - 1}}dx} \right\} \right)
    \nonumber 
 \end{align}
 \begin{align} \label{Eq:Apn01:02}
    & \mathop = \limits^{\left( b \right)} \exp \! \left( \!  - \pi \lambda _{{\rm{GW}}}^{\left( {\rm{I}} \right),\widetilde k,s}{{\left( l / K_0 \right)}^\delta }{\mathbb{E}_h}\left\{ {\left( {_1{F_1}\left( { - \delta ,1 - \delta , - sh/l} \right) - 1} \right)} \! \right\} \! \right)
    \nonumber \\
    & \mathop  = \limits^{\left( c \right)} \exp \left(  - \pi \lambda _{{\rm{GW}}}^{\left( {\rm{I}} \right),\widetilde k,s}{{\left( l /K_0 \right)}^\delta } \Theta \left( s/l \right) \right),
\end{align}
where $\mathbb{E}\left\{ . \right\}$ is the expectation operator; $\left( a \right)$ employs the probability generating functional
(PGFL) of PPP with density ${\lambda _{{\rm{GW}}}^{\left( {\rm{I}} \right),\widetilde k,s}}$; $\left( b \right)$ utilities following result: given $v \in (0,1)$, $a > 0$ and $b \in \mathbb{R}$, we have $\int\nolimits_a^\infty \left( {\exp \left( {b/x} \right) - 1} \right){x^{v - 1}}dx = \left( {1/v} \right){a^v}\left( {1 - {\;_1}{F_1}\left( { - v,1 - v,b/a} \right)} \right)$; and $\left( c \right)$ is held by following: given $a,b,c \in \mathbb{R}$, $\alpha \in \mathbb{N}$, we have $\int_0^\infty  {} {x^{\alpha  - 1}}\exp \left( { - cx} \right){\;_1}{F_1}\left( {a,b, - x} \right)dx = {c^{ - \alpha }}\Gamma \left( \alpha  \right){\;_2}{F_1}\left( {\alpha ,a,b, - 1/c} \right)$;
${\;_1}{F_1}\left( . \right)$ is the confluent hypergeometric function.
By substituting \eqref{Eq:Apn01:02} and the probability density function (PDF) of $L_0$, i.e., ${f_{{L_0}}}\left( l \right) = \delta \pi \lambda _{{\rm{GW}}}^{\left( {\rm{A}} \right)}{\left( {{K_0}} \right)^{ - \delta }}{l^{\delta  - 1}}\exp \left( { - \pi \lambda _{{\rm{GW}}}^{\left( {\rm{A}} \right)}{{\left( {l/{K_0}} \right)}^\delta }} \right)$ \cite{Thanh:EE_Cellular:01}, \eqref{Eq:Apn01:01} is then evaluated as 
{
\begin{align}
    & P_{S{\rm{IR}}}\left( {k,\widetilde k}, s \right) = \delta \pi \lambda _{{\rm{GW}}}^{\left( {\rm{A}} \right)}{\left( K_0 \right)^{-\delta} }\int\nolimits_{l = 0}^\infty 
    \exp \! \left( \! { - \pi \lambda _{{\rm{GW}}}^{\left( {\rm{A}} \right)}{{\left( l/K_0 \right)}^\delta }} \right)
    \nonumber \\
    & \times {l^{\delta  - 1}} \exp \left( { - \pi \lambda _{{\rm{GW}}}^{\left( {\rm{I}} \right),\widetilde k,s}{{\left( l / K_0 \right)}^\delta }\Theta \left( {{\Delta _{k,\widetilde k}}} \right)} \right) dl
    \nonumber \\
    & \mathop  = \limits^{\left( a \right)} \pi \lambda _{{\rm{GW}}}^{\left( {\rm{A}} \right)}{\left( 1/ K_0 \right)^\delta }
    \int\nolimits_{t = 0}^\infty \exp 
    \bigg(
    - \pi \lambda _{{\rm{GW}}}^{\left( {\rm{A}} \right)}t{{\left( 1/ K_0 \right)}^\delta }
    \bigg( 1 +
    \bigg. \bigg.
    \nonumber \\
    & \times
    \left. \left. \!
     p_{\widetilde k}^s{P_{{\rm{Act}}}}\Theta \left( {{\Delta _{k,\widetilde k}}} \right) \right) 
    \right)dt
    \! = \! {\left( {1 \! + p_{\widetilde k}^s{P_{{\rm{Act}}}}\Theta \left( {{\Delta _{k,\widetilde k}}} \! \right)} \! \right)^{- 1}}\!,
\end{align}
}
where $\left( a \right)$ follows from the change of variable $t = l^\delta$ and employing $\lambda _{{\rm{GW}}}^{\left( {\rm{I}} \right),\widetilde k,s} = p_{\widetilde k}^s{P_{{\rm{Act}}}}\lambda _{{\rm{GW}}}^{\left( {\rm{A}} \right)}$ QED.
%


\vspace{-0.25cm}
\begin{thebibliography}{00}
\vspace{-0.15cm}
%
{
%
\bibitem{EricReport:01} Ericsson Mobility Report, Ericsson, 2017.
%
\bibitem{Claire:01} C. Goursaud and J. M. Gorce, ``Dedicated networks for IoT: PHY/MAC state of theart and challenges", \emph{EAI Trans. IoT}, vol. 1, no. 1, 2015.
%
\bibitem{LoRa:DL_constraint:01} V. Di Vincenzo \textit{et al.}, "Improving Downlink Scalability in LoRaWAN," ICC 2019, Shanghai, China, 2019, pp. 1-7. 
%
\bibitem{Maalel:LoRa:01} N. Maalel \textit{et al.}, ``Reliability for Emergency Applications in Internet of Things," \emph{IEEE ICDCSS 2013}, 2013, pp. 361-366. 
%
\bibitem{LoRa:analysis:seminal:01}  O. Georgiou and U. Raza, ``Low Power Wide Area Network Analysis: Can LoRa Scale?", \emph{IEEE Wireless Commun. Lett.}, 2017.
%
\bibitem{SG_Matern:01}  Z. Qin \textit{et al.}, ``Performance Analysis of Clustered LoRa Networks", \emph{ IEEE Trans. Veh. Technol.}, vol. 68, no. 8, pp. 7616 - 7629, Aug. 2019. 
%
\bibitem{Beltramelli:ALOHA:01} L. Beltramelli \textit{et al.}, ``LoRa beyond ALOHA: An Investigation of Alternative Random Access Protocols," in \emph{IEEE Trans. Ind. Informat.}, vol. 17, no. 5, pp. 3544-3554, May 2021.
%
\bibitem{Reynders:faire-sf:01} B. Reynders \textit{et. al.}, ``Power and spreading factor control in low power wide area networks," \emph{IEE ICC 2017}, Paris, 2017, pp. 1-6.
%
%
\bibitem{Centenaro:01} M. Centenaro \textit{et. al.}, ``On the impact of downlink feedback on LoRa performance," \emph{IEEE PIMRC 2017}, Montreal, 2017. pp. 1-6.
%
\bibitem{Ron:01} D. Ron \textit{et. al.}, ``Performance Analysis and Optimization of Downlink Transmission in LoRaWAN Class B Mode," in \emph{IEEE Internet Things J.}, vol. 7, no. 8, pp. 7836-7847, Aug. 2020.
%

\bibitem{Andrews:SG:01} J. G. Andrews \textit{et. al.}, ``A Tractable Approach to Coverage and Rate in Cellular Networks," \emph{IEEE Trans. Commun.}, vol. 59, no. 11, 2011.
%
\bibitem{Tech_Report_ETSI:01} Tech. Rep. EN300 4.1, European Telecommunications Standards Institute, 2013. 
%
%
\bibitem{Thanh:LoRa_ICC:01} L.-T. Tu \textit{et. al.}, ``A New Closed-Form Expression of the Coverage Probability for Different QoS in LoRa Networks," \emph{IEE ICC 2020}.
%
\bibitem{Baccelli:book:01} F. Baccelli and B. Blaszczyszyn, Stochastic Geometry and Wireless Networks, Part I: Theory, Now Publishers, Sep. 2009.
%
\bibitem{Thanh:EE_Cellular:01} T. T. Lam \textit{et. al.}, ``On the Energy Efficiency of Heterogeneous Cellular Networks With Renewable Energy Sources—A Stochastic Geometry Framework," \emph{IEEE Trans. Wireless Commun.}, 2020.
%
\bibitem{Yu:VoronoiCell:01} S. M. Yu and S. Kim, ``Downlink capacity and base station density in cellular networks," ÌEEE WiOpt 2013, pp. 119-124.
%
\bibitem{Croce:01} D. Croce \textit{et. al.}, ``LoRa Technology Demystified:  from Link Behavior to Cell Capacity," \emph{IEEE Trans Wireless Commun.}, vol. 19, 2020.
%





%

%
%
%




}
\balance
%
\end{thebibliography}
\end{document}